\documentclass[12pt]{amsart}

\usepackage{amssymb,array,epsfig,color}

\newtheorem{lemma}{Lemma}

\newtheorem{theorem}[lemma]{Theorem}

\newcommand{\N}{{\mathbb N}}

\newcommand{\R}{{\mathbb R}}

\newcommand{\A}{\mathbf{A}}
\newcommand{\B}{\mathbf{B}}

\newcommand{\ud}{\mathrm{d}}

\begin{document}

\date{15th July 2007}

\title[The Supermembrane with Central Charges]
{The supermembrane with central charges:\\ (2+1)-D NCSYM,
confinement \\ and phase transition}
\author[L Boulton, M P Garcia del Moral and A Restuccia]
{L Boulton$^1$, M P Garcia del Moral$^2$ \\
and A Restuccia$^3$}

\begin{abstract}
The spectrum of the bosonic sector of the $D=11$ supermembrane with
central charges is shown to be discrete and with finite multiplicities,
hence containing a mass gap. The result extends to the exact theory
our previous proof of the similar property for the $SU(N)$ regularised
model and strongly suggest
discreteness of the spectrum for the complete Hamiltonian
of the supermembrane with central charges. This theory is a quantum
equivalent to a symplectic non-commutative super-Yang-Mills 
in $2+1$ dimensions, where the space-like sector is a Riemann surface of
positive genus. In this context, it is argued how
the theory in $4D$ exhibits confinement in the $N=1$ supermembrane with
central charges phase and how the theory enters in the
quark-gluon plasma phase through the spontaneous breaking of the
centre.  This phase is interpreted in terms of the compactified supermembrane without central charges.
\end{abstract}

\maketitle

\newpage

\section{Introduction} \label{s1}

\subsection{Motivation} \label{s1.1}
An important step towards a better understanding of the non-pertur\-bative
approach to  superstring theory is the non-perturbative
treatment of $D=11$ supermembranes \cite{bst}. The quantisation
of the latter, when it is embedded on Minkowski space-time, was studied
in \cite{dwln,dwmn,dwhn,hoppe} in terms of a quantum mechanical
maximally supersymmetric $SU(N)$ Yang-Mills matrix models. This was also
considered in a different context in \cite{halpern}.  In the seminal
work \cite{dwln},  it was shown that the spectrum of the $SU(N)$
regularised supersymmetric Hamiltonian is continuous, consisting of
the interval $[0,\infty)$. Remarkably, the spectrum
of the corresponding bosonic
Hamiltonian, equivalent to the dimensional reduction of
$D=10$ super-Yang-Mills to $0+1$
space-time, is discrete \cite{lucher,amilcar}. However
its configuration space comprises singular
string-like spikes, which, along with  supersymmetry, render the
spectrum continuous.

The validity of the
$SU(N)$ regularisation is justified by
the fact that the structure constants of the area-preserving
diffeomorphisms, the gauge symmetry of the supermembrane in the
light cone gauge, are equal to the large $N$ limit of the
$SU(N)$ structure constants. The characterisation of the spectrum
was performed in the $SU(N)$ regularised model, but we are not aware of
any result concerning
the large $N$ limit of the spectrum.

The supermembrane was interpreted
as a many-body object.  In this interpretation,
 the string-like spikes may connect
different membranes without changing the energy of the system,
in distinction to the standard situation in  string  theory. For a
review see \cite{helling}. The supermembrane embedded on a target space with a compact sector
was analysed in \cite{dwpp}. Although an $SU(N)$ regularisation was
not obtained, it was argued that the same qualitative features of
the spectrum remain valid.

The $D=11$ supermembrane with non-trivial central charges was
introduced in \cite{torrealba}. The configuration space of this model
is restricted by a topological condition. This
restriction implies the existence of a non-trivial central charge on
the SUSY algebra of the supermembrane. From a geometrical point of
view, the topological condition determines a non-trivial $U(1)$
principal bundle over the worldvolume whose canonical connections,
$U(1)$ monopoles, define minimal immersions into the compactified
sector of the target space. These immersions describe the wrapping
of the supermembrane on a calibrated submanifold of  the target
space. An interesting case  is $M_{4}\times T^{d}\times (S^{1})^{7-2g}$, where $M_{4}$ represents a Minkowski space-time in the light-cone gauge and $g=1,2,3$ 
is the genus of the 2-torus. See \cite{br} for the case $M_{7}\times T^{4}$.  The resulting theory preserves an $N=1$ supersymmetry in those dimensions.

The supermembrane with non-trivial central charges
does not contain string-like spikes and it admits an $SU(N)$
regularisation \cite{gmr}. The bosonic potential of this model increases
towards infinity as we move away from zero in the configuration space,
ensuring
a compact resolvent for the bosonic Hamiltonian \cite{bgmmr}. The
spectrum of the regularised Hamiltonian is discrete, with finite
multiplicity \cite{bgmr} and its heat kernel can be defined
rigorously by a process described in \cite{l1}.

For a $g\geq 2$ base manifold, this supermembrane  admits
an interpretation in terms of an orthogonal intersection of a
suitable number of genus 1 supermembranes with non-trivial central charge
\cite{br}.  In the type $IIA$ picture,  the theory compactified 
on a $T^{2}$, i.e. $g=1$, may be viewed as a
bundle of $D2-D0$ bound state theories where $D0$ monopole charges are
induced by non-constant fluxes on the $D2$ \cite{d2-d0}. Extensions
to $SU(N)$ interacting supermembranes may be considered as
in \cite{rita}.

In the cases where a minimal
immersion  from the base manifold
into the compact sector of the target space
can be established
(the former is a Riemann surface of genus $g$), the topological
restriction can be solved, and the supermembrane with non-trivial
central charges is equivalent (as a quantum field theory) to a
symplectic non-commutative super-Yang-Mills  theory (NCSYM)
\cite{ovalle, mr}.
This is the case, for instance, when the base manifold is a genus $g$
Riemann surface and the compact sector of the target space is a flat
$T^{2g}$ torus. The symplectic structure is determined by the
minimal immersion, and describes the curvature of $U(1)$ monopole
connections.

The BFSS conjecture takes the $D0$ action as a fundamental action
\cite{BFSS}. This is precisely the same Hamiltonian than the SU(N)
regularised Hamiltonian of the $D=11$ supermembrane in the light
cone gauge \cite{dwln}. Its extension to interesting compactified
target spaces as the BMN model \cite{bmn} includes Chern-Simons and
mass terms, preserving the PP wave supersymmetry, allowing the existence
of stable vacum solutions. In the same sense, the topological condition
on the $D=11$ supermembrane with central charges give rise to mass terms which
render the theory stable. However, the topological properties
of the base manifold plays a fundamental role in the large $N$ limit
analysis of the regularised supermembrane. This geometrical structure
is missing in the matrix model, where no rigorous analysis
of the large $N$ limit has been obtained.

Non-commutative
Yang-Mills theories have been considered as toy models for gravity
\cite{gross} and \cite{szabo}.
The relation between supergravity and non-commutative
Yang-Mills become natural in the context of supermembranes, since
they are embedded on a target space which must be a solution of
$D=11$ supergravity. Moreover the supermultiplet of $D=11$
supergravity has been conjectured to be the ground state of the
theory.

In the context of  string  theory, NCSYM appears in a very
natural manner by wrapping D-branes \cite{cornalba-schiappa}.
A super-Yang-Mills
theory on a non-commutative torus is naturally related to the
compactification of a matrix theory on a dual torus with a constant
$C_{3}$ field, see for example,
\cite{douglas-connes,miao,douglas-hull,ishibashi}.
The non-commutative Yang-Mills theory
in a flat space with a rational non-commutative parameter
is related to ordinary Yang-Mills theories with magnetic flux
through Morita equivalence, \cite{bigatti}. By comparing ordinary
Yang-Mills theories and non-commutative ones, it was
found in \cite{maldacena-ruso} that both theories share the same
degrees of freedom in the IR limit.
However these degrees are redistributed
differently in both theories in the UV limit.

It was argued in \cite{ishibashi,hashimoto} that it should
be the non-commutative theory
the better suited  for describing the IR limit and the
commutative theory the one that best recovers the UV limit.

Yang-Mills theories with boundary modelling QCD where extensively studied
under the common name of ``bag models'' \cite{chodos}. 
 super-Yang-Mills theories behave differently at extremal 
energies since they are either in the confined or in the screening phase. 
For the bosonic theory, the confined phase occurs at
low temperatures where vector-like
gluons form  singlet  bound states of colour called
glueballs, see \cite{pepe}. 

The low temperature regime is characterised by the dynamics of gluons.
At high temperatures the gluons do no give rise to bound states but
rather to a plasma that constitute the screening phase. 
If the fermions are
introduced in the theory, at low energies
they feel a binding force
and at high energy  they remain free, \cite{thooft}. 
These two extremal regimes are thought in general to be
separated by a phase transition occurring when a global symmetry
of the theory breaks. This phase transition
is related to the spontaneous breaking
of the centre, \cite{thooft,thooft2}. Furthermore,
according to \cite{thooft,thooft2}, 
this symmetry is conjectured to be of a topological nature 
due to magnetic monopoles or instantons. 

The role of the centre has been considered to a large extent,
cf. e.g. \cite{pepe, aharony,pepe2}. 
According to the results of \cite{witten}, 
the instanton gas picture is only appropriate in those cases where
the topological charge is discrete. Otherwise the correct picture would
be the monopole. It is this latter case, where 
the confinement should appear, \cite{witten}, as it is 
for $D=4$ and $N=1$ in the supermembrane 
with central charges. 

Connections between membrane 
theory and Yang-Mills theories have been considered
in a different context in \cite{gabadadze}. In \cite{yaffe} 
the critical behaviour of a gauge theory in the de-confined phase
is related with the behaviour of a scalar which has a symmetry 
induced by the
centre of the group. The phase transition occurs when the
topological defect is metastable and decays
as a consequence  of quantum
processes.

\subsection{Aims and scopes of the present paper} \label{s1.2}
Our main goal is to
demonstrate that the bosonic Hamiltonian of the
supermembrane with non-trivial central charges has a discrete
spectrum with finite multiplicity.
Based on this observation we will show strong evidence
confirming that the spectrum of the supersymmetric
Hamiltonian possesses the same qualitative properties.
Consequently, the $N=1$ NCSYM theory in $2+1$ D is  equivalent to the
supermembrane with non-trivial central charge  has the same spectral
properties.

We also prove that, in the semiclassical regime, 
the large $N$ limit of
the eigenvalues of the regularised Hamiltonian
converge to those of the supermembrane with central charge.
The spectrum of the Hamiltonian exhibits a mass gap. The scalar
fields acquire a mass induced by the central charge of the
SUSY algebra. One may also argue that this mass is induced by the
centre $Z(2)$ of the
symplectic group associated to the central charge.  Supermembranes
with central charge and $N=1$ in $4$D may then be interpreted as the IR phase of a
more general theory. This is achieved by increasing the energy the centre breaks
spontaneously, so that the transition phase occurs in a screening
phase. This should correspond to an $N=4$ compactified supermembrane without central charges.
\medskip

In Section~\ref{s2}, we describe the
Hamiltonian of the supermembrane with central charges
and its semiclassical regime.  In
Section~\ref{s3} we establish the main results of the present paper, 
a crucial operator bounds connecting the
exact bosonic Hamiltonian and the semiclassical Hamiltonian. 
This operator bound, see \eqref{b1}, allows us to deduce spectral
properties of the exact bosonic Hamiltonian from those of the 
semiclassical one discussed in Section~\ref{s2}.
Natural questions arises concerning the large $N$ limit of the regularised
Hamiltonian.
We address these questions in Section~\ref{s4}. 
In the final Section~\ref{s6} 
we consider confinement properties of the theory in terms
of the centre of the group both at the exact and the regularised level.
We also discuss
the transition phase to de-confinement and give an interpretation in
terms of the supermembrane. An appendix is included, where
we explicitly compute the $SU(N)$ gauge symmetry in the regularisation model.


\section{The supermembranes with central charges} \label{s2}

In this section we analyse the
exact  action of the supermembrane with central charges
as well as its semiclassical approximation.

 Let the $D=11$
supermembrane be defined in terms of a base manifold, a $g=1$
Riemann surface $\Sigma$, and a target space $M_9\times S^1 \times
S^1$. Consider its formulation in the light cone gauge where the
directions $X^+,\,X^-,\,P_+$ and $P_-$ have been removed in the
standard manner \cite{dwhn}. The canonically reduced Hamiltonian has
the expression
\begin{equation*}
   \int_\Sigma  \sqrt{W} \left(\frac{1}{2}
\left(\frac{P_M}{\sqrt{W}}\right)^2 +\frac{1}{4} \{X^M,X^N\}^2+
({\small\mathrm{fermionic\ terms}})\right)
\end{equation*}
subject to the condition
\begin{equation} \label{e1}
   \oint_\mathcal{C} \frac{P_M}{\sqrt{W}} \ud X^M = 0.
\end{equation}
Here and below $M,N=1,\ldots,9$. The integral in
(\ref{e1})  is the generator of an area preserving diffeomorphism
of $\Sigma$ for $\mathcal{C}$ any given closed path. This
constraint may be expressed as a local condition
\begin{equation*}
    \{P_M,X^M\} \equiv \epsilon^{ab} \partial_a\left(\frac{P_M}{\sqrt{W}}
\right) \partial_b X^M =0,
\end{equation*}
which generates area preserving diffeomorphisms associated to exact
one-forms coupled to the global constraint
\begin{equation} \label{f1}
   \oint _{C_i}\frac{P_M}{\sqrt{W}} \ud X^M =0, \qquad i=1,2,
\end{equation}
where $C_1$ and $C_2$ form of a basis of homology on $\Sigma$
which generates area preserving diffeomorphisms associated to harmonic
one-forms.

The scalar density $\sqrt{W}$ is present in expression (\ref{f1})
 as a consequence
of the gauge fixing procedure and it is preserved by the diffeomorphisms
mentioned above. Let us now impose some topological restrictions
on the configuration space which completely characterise the
$D=11$ supermembrane with non-trivial central charge generated by
the wrapping on the compact sector of the target space. All maps
from the base space $\Sigma$, must satisfy
\begin{equation} \label{e2}
\begin{aligned}
  &\oint_{C_i} \ud X^r=2\pi S^r_iR^r, \qquad &r=1,2 , \\
  &\oint_{C_i} \ud X^m=0 \qquad &m=3,\ldots,9,
\end{aligned}
\end{equation}
for $i=1,2$ and
\begin{equation} \label{e3}
   \int_\Sigma \ud X^r\wedge\ud X^s =\epsilon^{rs}(2\pi^2R_1R_2)n,
\end{equation}
where $n=\det S^{r}_i$ is fixed, each entry $S^r_i$ is integer,
and $R_1$ and $R_2$ denote the radii of the target component
$S^1\times S^1$. Note that (\ref{e2}) describe maps from $\Sigma$
to $S^1\times S^1$ with $\ud X^m$ exact one-forms and $\ud X^r$
non-trivial closed one-forms.
The only restriction upon these maps is the assumption that $n$ is
fixed. The term on the left side of (\ref{e3}) describes the
central charge of the supersymmetric algebra. As we shall see
next, the factor $R_1R_2(2\pi)^2$ is the area of $\Sigma$ in the
induced metric.

The general map satisfying (\ref{e2})-(\ref{e3}) can be constructed
explicitly. Any closed one-form $\ud X^r$ may be expressed as the sum
of a harmonic and an exact form,
\begin{equation}\label{ee}
  \ud X^r=L^r_s \ud \hat{X}^s + \delta_s^r \ud A_s \qquad s,r=1,2
\end{equation}
where $L^r_s$ are real numbers and $\ud \hat{X}$ is a canonical
basis of harmonic one-forms over $\Sigma$. The term $\ud
\hat{X}^s$, $s=1,2$, is found by considering the (unique)
holomorphic one-form $\omega$, normalised with respect to the
elements of the homology basis $C_i$, defined by:
\begin{equation*}
   \oint_{C_1}w=1, \qquad \oint_{C_2} w=\Pi,
\end{equation*}
where $\Pi$ is the period of $\omega$ in the basis given by $C_i$.
By construction, the imaginary part of $\Pi$ is positive. Let
\cite{br}
\begin{equation*}
   \omega=\ud \widetilde{X}^1+i \ud \widetilde{X}^2
\end{equation*}
and define
\begin{equation*}
   \ud \hat{X}^r = (M^{-1} \ud \tilde{X})^r,
\end{equation*}
where the constant matrix $M$ is given by
\begin{equation*}
     M=\begin{pmatrix} 1 & \mathrm{Re}\,\Pi \\
                       0 & \mathrm{Im}\, \Pi \end{pmatrix}.
\end{equation*}
Then \cite{br}
\begin{equation*}
    \oint_{C_i} \ud \hat{X}^r=\delta_i^r
\end{equation*}
and
\begin{equation*}
    \int_\Sigma \ud \hat{X}^r \wedge \ud \hat{X}^s =\epsilon^{rs}.
\end{equation*}
If (\ref{e2}) is to be satisfied, necessarily
\begin{equation*}
   L^r_s=2\pi R^rS^r_s.
\end{equation*}
Condition (\ref{e3}) implies
\begin{equation*}
   S^r_s S^t_u \epsilon ^{su} = n \epsilon ^{rt}.
\end{equation*}
Define the scalar density $\sqrt{W}$ by
\begin{equation*}
   \epsilon^{ab} \partial_a \hat{X}^r \partial_b \hat{X}^s\epsilon_{rs}=
\sqrt{W}
\end{equation*}
where $\partial_a\equiv \partial/\partial\sigma^a$, $a=1,2$, $\sigma^a$
being local coordinates on $\Sigma$. Then
\begin{equation*}
   \epsilon_{rs} \ \ud \hat{X}^r\wedge \ud \hat{X}^s=
   \sqrt{W}\ \ud \sigma^1 \wedge \ud \sigma^2.
\end{equation*}
A change of the canonical basis of homology over $\Sigma$, implies
varying the corresponding harmonic one-form $\ud \hat{X}^r
\leadsto T^r_s \ud \hat{X}^s$, where $T\in SL(2,Z)$, that is
\begin{equation*}
    T^r_sT^t_u \epsilon^{su}=\epsilon^{rt},
\end{equation*}
$T^r_s$ integers. As the density $\sqrt{W}$ remains invariant under
these transformations, they are area-preserving diffeomorphisms
disconnected from the identity. The theory is then invariant under
$SL(2,Z)$. The degrees of freedom are expressed in terms of
$A_{r}$ and the discrete set of integers described by the harmonic
one-forms. We can always fix these gauge transformations, within the conformal class of the target torus, by
\begin{equation*}
   S^r_i=l^r \delta^r_i, \qquad l^1 l^2=n.
\end{equation*}
Therefore
\begin{equation*}
   \ud X^r=2\pi R^r l^r \ud \hat{X}^r + \delta^r_s \ud A_s.
\end{equation*}
After the gauge fixing there is a residual invariance
$\mathbb{Z}(2)$.

 The complete expression for the
Hamiltonian of the $D=11$ supermembrane subject to the topological
conditions (\ref{e2}) and (\ref{e3}) turns out to be \cite{bgmr,ovalle,gmr},
\begin{equation} \label{e4}
\begin{aligned}
   H= \int \sqrt{W} \ud \sigma^1\wedge \ud\sigma^2
   \left[\frac{1}{2} \left(\frac{P_m}{\sqrt{W}} \right)^2
   +\frac{1}{2} \left(\frac{\Pi^r}{\sqrt{W}} \right)^2
   + \frac{1}{4} \{X^m,X^n\}^2 + \right.\\ \left.
   \frac{1}{2}(\mathcal{D}_rX^m)^2
   +\frac{1}{4}(\mathcal{F}_{rs})^2 +\Lambda(\{P_m,X^m\}
   + \mathcal{D}_r\Pi^r ) \right]+
   (\mathrm{fermionic\ term}).
\end{aligned}
\end{equation}
where \cite{ovalle,gmr}
\begin{equation*}
\begin{aligned}
 \mathcal{D}_rX^m=D_rX^m+\{A_r,X^m\}, \\
   \mathcal{D}_r\Pi^r=2\pi R^rl^r \frac{\epsilon^{ab}}{\sqrt{W}}
   \partial_a \hat{X}^r \partial_b \left(\frac{\Pi^r}{\sqrt{W}}\right)+
[A_{r},\Pi^{r}],\\
   \mathcal{F}_{rs}=D_rA_s-D_s A_r+ [A_r,A_s]
\end{aligned}
\end{equation*}
\vspace{0.5cm}
and $D_r=2\pi R^rl^r \frac{\epsilon^{ab}}{\sqrt{W}}\partial_a
\hat{X}^r \partial_b$.  The associated mass operator is
$(\mathrm{mass})^{2}=Z^{2}+H$,
where $Z^2=\frac{1}{8}((2\pi)^2nR_{1}R_{2})^{2}$ is the central charge.

\subsection{Semiclassical regime of the bosonic Hamiltonian}

The semiclassical approximation of the theory is
obtained by considering only the quadratic terms in the above
expression for the Hamiltonian. Let
\begin{equation*}
\begin{aligned}
   H_{\mathrm{sc}} = \int \sqrt{W} \ud \sigma^1\wedge \ud\sigma^2
   \left[\frac{1}{2} \left(\frac{P_m}{\sqrt{W}} \right)^2
   +\frac{1}{2} \left(\frac{\Pi^r}{\sqrt{W}} \right)^2
   + \right.\\ \left.
   \frac{1}{2}(D_rX^m)^2
   +\frac{1}{4}(\widehat{\mathcal{F}}_{rs})^2 +
   \Lambda D_r\Pi^r \right]+
   (\mathrm{fermionic\ terms}),
\end{aligned}
\end{equation*}
where, in the semiclassical approximation,
\begin{gather}
   \widehat{\mathcal{F}}_{rs}=D_rA_s-D_s A_r.
\end{gather}
The general solution to the constraint $D_r\Pi^r=0$ is
\begin{equation*}
   \Pi^r = \epsilon^{rs} 2\pi R^sl^s \epsilon^{ab} \partial_a \hat{X}^s
   \partial _b \left( \frac{\Pi}{\sqrt{W}} \right),
\end{equation*}
where $\Pi$ is a scalar density.

The kinetic term $\Pi^r\dot{A}_r$  may be rewritten, once integration
by parts has been carried out, as $p\dot{q}$ where $p=\sqrt{W}\frac{1}{2}\epsilon ^{rs}
\mathcal{F}_{rs}$ and $q=\frac{\Pi}{\sqrt{W}}$. Thus
\begin{equation*}
\begin{aligned}
   H=\int \ud \sigma^1 \wedge \ud \sigma^2
   \sqrt{W}\left[\frac{1}{2}\left(\frac{\Pi^r}{\sqrt{W}} \right)^2+
   \frac{1}{4}(\widehat{\mathcal{F}}_{rs})^2+\Lambda D_r\Pi^r\right]=\\
   =\frac{1}{k}\int \ud \sigma^1 \wedge \ud \sigma^2
   \sqrt{W}\left[\frac{1}{2}\left(\frac{p}{\sqrt{W}} \right)^2+
   \frac{1}{2}(D_r q)^2\right]
\end{aligned}
\end{equation*}
which coincides with the contribution to the Hamiltonian
of the transverse modes $X^m$, $m=3,\ldots,9$.

The above shows that, from a gauge independent point of view, the
complete bosonic Hamiltonian in the semiclassical approximation
is
\begin{equation*}
   H_{\mathrm{sc}}^B =\int
    \ud \sigma^1 \wedge \ud \sigma^2
   \sqrt{W}\left[\frac{1}{2}\left(\frac{P^M}{\sqrt{W}} \right)^2+
   \frac{1}{2}(D_r X^M)^2\right]
\end{equation*}
where $M=1,\ldots,8$. If we now express $X^M$ and $P^M/\sqrt{W}$
in terms of a complete orthonormal basis of scalar symmetries over
over $\Sigma$, we obtain
\begin{gather}
    X^M=X^M_A (\tau)\exp[2\pi i (a_r\hat{X}^r)](\sigma), \\
    \frac{P^M}{\sqrt{W}}=\rho^M_A(\tau) \exp[2\pi i(a_r\hat{X}^r)](\sigma),
\end{gather}
where $A=(a_1,a_2)$. Thus, the bosonic contribution in the
semiclassical Hamiltonian takes the form
\begin{equation*}
   H^\mathrm{B}_{\mathrm{sc}}=\frac{1}{2}[(\rho^M_A)^2+
  \omega_A^2(X^m_A)^2].
\end{equation*}

The spectrum of $H^\mathrm{B}_{\mathrm{sc}}$ is then characterised
in the following fashion. For any finite subset $\Omega$ of
$\N\times \N$ there is an eigenvalue
\begin{equation} \label{f2}
\begin{gathered}
   \lambda_\Omega =\sum_{A\in\Omega} \omega_A,  \\
    \omega_A=\pi^2 \sqrt{(R^1l^1a_2)^2+(R^2 l^2a_1)^2}.
\end{gathered}
\end{equation}
This expression coincides with the particular case considered in
\cite{stelle} where the semiclassical quantization of the supermembrane over a fixed background
 was considered. The model in \cite{stelle} correspond to a gauge fixed version of the supermembrane with central charges in the semiclassical 
approximation.  By virtue of \eqref{f2}, for any given energy level
$E$, there only exists a finite number of eigenvalues of
$H^\mathrm{B}_\mathrm{sc}$ below $E$. The spectrum is therefore discrete.

This is the expression of the eigenvalues when the zero point energy
has been eliminated. It is automatically cancelled when the
semiclassical supersymmetric Hamiltonian is considered. This
property was first demonstrated in \cite{stelle} and it coincides
exactly with
the one for the semiclassical supermembrane with central charges.


\section{Operator bounds on the exact bosonic Hamiltonian}
\label{s3}

According to the results reported in \cite{bgmmr}, the bosonic
regularised Hamiltonian of the $D=11$ supermembrane with central
charge, $H_{N}^{B}$, relates to its semiclassical approximation,
$H_{\mathrm{sc},N}^{B}$, by means of the following operator
inequality:
\begin{align} \label{fi3}
H_{N}^{B}\geq C_{N}H_{\mathrm{sc},N}^{B}.
\end{align}
Here $N$ denotes the size of the truncation in the Fourier basis of
$\Sigma$ and $C_N$ is a positive constant dependent on the size of the
truncation. A  crucial step
in the proof of \eqref{fi3}, relies heavily on
the compactness of the unit ball of the configuration space which
happens to be finite dimensional. In this section we show that the
same operator relation holds true for the exact bosonic
Hamiltonians, i.e. the case $N=\infty$, see Theorem~\ref{t1}. The main source of difficulties
in the proof of this assertion lies in the fact that now the unit
ball of the configuration space does not possess the property of
being compact. We overcome these difficulties by carrying out a
detailed analysis of each term involved in the expansion of the
potential term of $H^B$.

\subsection{The configuration space and the gauge fixing condition}
\label{s5.1}

We firstly construct the configuration space
for the supermembrane with
central charges in a rigorous fashion
and prescribe suitable gauge fixing condition.

Constant functions are harmonic,
therefore the constant modes of the fields $X^{m}$ and $A_{r}$ belong
to the harmonic sector in \eqref{ee}. In addition only derivatives of
$X^m$ and $A_r$ appear in all the expressions.
  Let $\mathcal{H}^{1}$ denote
the Hilbert space obtained by completing $C^{1}(\Sigma)$, modulo
locally constant functions, with respect to the norm
\begin{align*}
\|u\|^{2}=\int \ud
^{2}\sigma\sqrt{W}g^{ab}\partial_{a}u\partial_{b}\overline{u},
\end{align*}
where $g^{ab}$ is the inverse of the metric
$g_{ab}=\partial_{a}\widehat{X}^{r}\partial_{b}\widehat{X}^{r}$
induced over $\Sigma$ by the minimal immersion $\widehat{X}_{r}$.
Below we employ the following convention: for a field $u$,
\[
    \langle u \rangle = \int_{\Sigma}\ud ^{2}\sigma\sqrt{W}\, u.
\]
For simplicity we will consider below only the case $n=1$ and $R_1=R_2=1$.

Note that
\begin{align*}
D_{r}uD_{r}\overline{u}&=
\frac{\epsilon^{ab}}{\sqrt{W}}\partial_{a}\widehat{X}^{r}\partial_{c}u
\frac{\epsilon^{bd}}{\sqrt{W}}\partial_{b}\widehat{X}^{r}\partial_{d}
\overline{u}\\
\nonumber &=g^{cd}\partial_{c}u\partial_{d}\overline{u}
\end{align*}
and
\begin{align*}
\{u,w\}\equiv\frac{\epsilon^{ab}}{\sqrt{w}}\partial_{a}u\partial_{b}w=
\epsilon^{rs}D_{r}uD_{s}w,
\end{align*}
so that
\begin{align*}
\|u\|^{2}=\langle D_{r}uD_{r}\overline{u} \rangle.
\end{align*}

Let
\begin{align} \label{fi7}
\|u\|_{4,2}=(\langle D_{r}u\rangle^2+\langle D_rD_s u\rangle^2)^{1/4}.
\end{align}
Below and elsewhere $X^m$, $A_r$ are assumed to be members of
the configuration space $\mathcal{H}^{4,2}$ of fields
$u\in \mathcal{H}^1$ such that $\|u\|_{4,2}<\infty$. The
left hand side of \eqref{fi7} is a well defined norm in
$\mathcal{H}^{4,2}$. The latter is a linear space, however we do not
make any assumption about completeness.

The potential, $V$, of the bosonic sector of the supermembrane
with central charges is well defined in $\mathcal{H}^{4,2}$ as
\begin{align} \label{fi1}
V=\langle
\mathcal{D}_{r}X^{m}\mathcal{D}_{r}X^{m}+\frac{1}{4}\mathcal{F}_{rs}
\mathcal{F}_{rs} \rangle.
\end{align}
The introduction of the constrained space $\mathcal{H}^{4,2}\subset
\mathcal{H}^1$ is
justified by the fact that homogeneous terms of order 4 are
present on the right hand side of \eqref{fi1}, so $V$ is not
well defined in the whole space $\mathcal{H}^1$.

\medskip

The following gauge fixing conditions are equivalent to those considered
in \cite{gmr, bgmmr, bgmr},
\begin{equation} \label{fi2}
\begin{aligned}
&D_{1}A_{1}=0\\
&\mathrm{if}\hspace{.3cm}D_{1}A_{2}=0 \hspace{.4cm}\mathrm{then}
\hspace{.3cm} A_{2}=0.
\end{aligned}
\end{equation}
These
are obtained by expressing the fields in terms of an orthonormal
basis over $\Sigma$. Integration by parts yields
$
\langle D_{2}A_{1}D_{1}A_{2} \rangle =0.
$
Similarly we also have
$
\langle D_{2}A_{1}\{A_{1},A_{2}\}\rangle =0.
$
Note that
$
\langle (D_{1}A_{2})^{2} \rangle=0,
$
implies $A_{2}=0$.

\subsection{The uniform quadratic bound for the bosonic potential}
\label{s5.2}
Let $\rho^{2}$ be the potential term of
$H^B_{\mathrm{sc}}$, so that
\begin{align*}
\rho^{2}=\langle D_{r}X^{m}D_{r}X^{m}+
(D_{1}A_{2})^{2}+(D_{2}A_{1})^{2} \rangle.
\end{align*}
We may rewrite
\begin{equation*}
V=\rho^{2}+2\B+\A^2
\end{equation*}
where
\begin{gather*}
\B=\langle D_rX^m\{A_r,X^m\}+D_1A_2\{A_1,A_2\} \rangle, \\
\A ^2=\langle \{A_1,X^m\}^2+\{A_2,X^m\}^2+\{A_1,A_2\}^2+\{X^m,X^n\}^2
\rangle.
\end{gather*}

\begin{theorem} \label{t1}
There exists a constant $0< C\leq 1$, such that
\begin{equation} \label{fi4}
  V\geq C \rho ^2, \qquad \qquad \forall X^m,\, A_r \in \mathcal{H}^{4,2}.
\end{equation}
\end{theorem}

The remaining parts of this section is devoted to showing the validity
of Theorem~\ref{t1}.

By definition,
\begin{equation*}
   V=\rho^2(1+\frac{2\B}{\rho^2}+\frac{\A^2}{\rho^2})
     = \rho^2(1+2bR+a^2R^2),
\end{equation*}
where
\begin{gather}
    R=\| (X^m,A_r)\|_{4,2}, \nonumber \\
    a^2=\frac{\A^2}{R^2\rho^2} \qquad\mathrm{and}
    \qquad b=\frac{\B}{\rho^2R}   \label{fi9}.
\end{gather}
Since both terms $a^2$ and $b$ are homogeneous in $X^m$ and $A_r$,
they are constant in $R$. Without loss of generality we assume
that $a^2$ and $b$ are always evaluated at fields $X^m$, $A_r$, normalised
by the condition $R=1$.

Let
\[
     P(R)=1+2bR+a^2R^2
\]
be the real polynomial whose variable is $R\geq 0$. The existence of a constant $C>0$ satisfying \eqref{fi4} is
equivalent to the condition
\begin{equation} \label{fi8}
  \inf _{\|(X^m,A_r)\|_{4,2}=1} \left [ \inf _{R\geq 0}
  P(R) \right] \geq C.
\end{equation}

Note that $\B$ is the inner product in $\mathcal{H}_1$ of the
field \[(D_1X^m,D_2X^m, D_1A_2, 0)\] and the field
\[(\{A_1,X^m\},\{A_2,X^m\},\{A_1,A_2\},\{X^m,X^n\}),\] while $\A$
is the norm of the latter. Thus $\A^2=0$ yields $\B=0$, so the
condition $a^2=0$ implies $P(R)=1$. Hence
\begin{align*}
    \inf _{\|(X^m,A_r)\|_{4,2}=1} \left [ \inf _{R\geq 0}
  P(R) \right] &= \min \left[ 1,
  \inf_{a\not=0} \left( 1-\frac{b^2}{a^2}\right)
  \right]\\
  &=\inf_{a\not=0}\left( 1-\frac{b^2}{a^2} \right)
\end{align*}
in \eqref{fi8}.

The validity of the following lemma will immediately ensure
\eqref{fi8}, hence Theorem~\ref{t1}.

\begin{lemma} \label{t2}
Let $a$ and $b$ be the quantities defined by \eqref{fi9}. Then
\[
   \inf_{a\not=0} \left( 1-\frac{b^2}{a^2}\right)>0.
\]
\end{lemma}
\proof We proceed by contradiction. Suppose that
\[
   \inf_{a\not=0} \left( 1-\frac{b^2}{a^2}\right)=0.
\]
Then we can find sequences $(X^m)_j$, $(A_r)_j\in \mathcal{H}^{4,2}$,
such that $\|((X^m)_j,(A_r)_j)\|_{4,2}=1$ and
\begin{equation} \label{fi10}
     \frac{b_j^2}{a_j^2}\to 1
\end{equation}
as $j\to \infty$, where $a_j,\, b_j$ are the quantities defined
by \eqref{fi9} for the fields at the $j$th place of the sequences.

By construction
\[
   \frac{b^2}{a^2}= \frac{\langle
   D_r X^m\{A_r,X^m\}+D_1 A_2\{A_1,A_2\}+(D_2 A_1) 0
   +0 \{X^m,X^n\} \rangle}{\rho^2 \A^2}^{2}.
\]
Thus, for each $j$, the left hand side of \eqref{fi10} is
the inner product of two unit vectors in
$\mathcal{H}^1$. By virtue of the Cauchy Schwarz inequality, these
two vectors should become increasingly parallel as as $j\to
\infty$. Since the quantities $a^2$ and $b^2$ remain constant if
we multiply the field $(X^m,A_r)$ by a constant, without loss of
generality we can chose our sequences $(X^m)_j$, $(A_r)_j$, such that
\begin{gather}
   \left\langle \left( \frac{D_r(X^m)_j}{\rho_j}-
   \frac{\{(A_r)_j,(X^m)_j\}}{\A_j}\right)^2 \right\rangle \to 0,
   \label{fi11} \\
  \left\langle \left( \frac{D_1(A_2)_j}{\rho_j}-
   \frac{\{(A_1)_j,(A_2)_j\}}{\A_j}\right)^2 \right\rangle \to 0,
   \label{fi12}\\
   \left\langle \left( \frac{D_2(A_1)_j}{\rho_j}\right)^2 \right\rangle \to 0
   \quad \mathrm{and} \quad
   \left\langle \left(
   \frac{\{(X^m)_j,(X^n)_j\}}{\A_j}\right)^2 \right\rangle
   \to 0.
\end{gather}

When $r=1$, the left side of \eqref{fi11}, is
\begin{align*}
 \left\langle \left( \frac{D_1(X^m)_j}{\rho_j}\right. \right. &
 \left. \left. -
   \frac{D_2 (A_1)_j D_1 (X^m)_j}{\A_j}\right)^2 \right\rangle
   \geq
   \\ & \left\langle \left( \frac{D_1(X^m)_j}{\rho_j} \right)^2
   \right \rangle
   - 2\left| \left\langle
   \frac{D_1 (X^m)_j}{\rho_j} \frac{D_2 (A_1)_j D_1 (X^m)_j}{\A_j}
   \right\rangle \right|.
\end{align*}
By virtue of the Cauchy Schwarz inequality,
\begin{align*}
   \left| \left\langle
   \frac{D_1 (X^m)_j}{\rho_j}\right. \right. &
   \left. \left. \frac{D_2 (A_1)_j D_1 (X^m)_j}{\A_j}
   \right\rangle \right| =
  \left| \left\langle
   (D_1 (X^m)_j)^2 \frac{D_2 (A_1)_j}{\rho_j \A_j}
   \right\rangle \right| \\
   & \leq \langle (D_1 (X^m)_j)^4 \rangle ^{1/2}
    \left\langle \left(\frac{D_2 (A_1)_j}{\rho_j}\right)^2
    \right\rangle ^{1/2} \frac{1}{\A_j} \\
    & \leq
    \left\langle \left(\frac{D_2 (A_1)_j}{\rho_j}\right)^2
    \right\rangle ^{1/2} \frac{1}{\A_j}.
\end{align*}
Furthermore, analogous results also hold true for the left sides of
\eqref{fi11} when $r=2$ and the left side of \eqref{fi12}. Hence, if $\A_j\not\to
0$, the above, along with \eqref{fi11} and \eqref{fi12}, imply
\[
    \frac{\langle (D_r (X^m)_j)^2+ (D_1 (A_2)_j)^2+(D_2(A_1)_j)^2
    \rangle}{\rho_j^2}\to 0
\]
which is impossible.

It is only left showing that the case $\A_j\to 0$ also leads to a
contradiction. We proceed as follows.

Let $\Delta$ denote the Laplacian operator acting on
$L^2(\Sigma)$. Integration by parts yields
\begin{align*}
   \langle D_r(D_1 (X^m)_j)^2 D_r (D_1 (X^m)_j)^2 \rangle &
   =  \langle [(-\Delta)^{1/2}  (D_1(X^m)_j)^2]^2 \rangle.
\end{align*}
Then, $(W^m)_j=(-\Delta)^{1/2}  (D_1(X^m)_j)^2 \in L^2(\Sigma)$.
Since $\|(X^m)_j\|_{4,2}\leq 1$, $\|(W^m)_j\|_{L^2}\leq 1$. Now
$(-\Delta)^{-1/2}$ is a compact operator and
\[(D_1(X^m)_j)^2=(-\Delta)^{-1/2}(W^m)_j.\] Thus $(D_1(X^m)_j)^2$
has a subsequence which is convergent in $\|\cdot\|_{L^2}$ to an
accumulation point, say $Y^m_1\in L^2(\Sigma)$.

Similarly $(D_2(X^m)_j)^2$, $(D_1(A_2)_j)^2$ and $(D_2(A_1)_j)^2$
have corresponding subsequences, convergent in $\|\cdot\|_{L^2}$ to
accumulation points, $Y^m_2$, $Z_1$ and $Z_2$ in $L^2(\Sigma)$.
Furthermore $Y^m_r$ and $Z_r$ lie on $L^4(\Sigma)$, so that we can
evaluate $P(R)$ at $((Y^m_r)^{1/2},(Z_r)^{1/2})$. As $\A^2=0$  when we
evaluate at these limit fields, in fact $P(R)$ achieves the constant
value 1. But, since $b$ and $a^2$ are continuous in
$(D_rX^m,D_1A_2,D_2A_1)$ for the norm $\|\cdot\|_{L^2}$, this
contradicts the condition $b_j^2/a_j^2\to 1$. The
proof of the lemma is hence completed.

\bigskip

In order to define rigorously the bosonic Hamiltonian in the non-compact
infinite dimensional configuration space $\mathcal{H}^{4,2}$, the operator  is expressed as
\begin{align*}
  H^{B}=[V_{\mathrm{quartic}}+V_{\mathrm{cubic}}+(1-C)V_{\mathrm{quadratic}}]+[-\Delta+C V_{\mathrm{quadratic}}].
\end{align*}
The term inside the first bracket acts as a multiplication operator on the Hilbert space
of states. The one inside the second bracket may be expressed
in terms of creation and annihilation operators.

By virtue of Theorem~\ref{t1},
\begin{equation} \label{b1}
  H^{B}\ge-\Delta+C V_{\mathrm{quadratic}}.
\end{equation}
The infinite zero point
energy can be extracted from the same operator at both sides of this inequality. This
zero point energy will be automatically cancelled out when we considered
the supersymmetric theory. According to the results
of \cite{bgmr} and \cite{l1}, the fermionic contribution does not
change the qualitatively properties of the bosonic Hamiltonian
of the regularised model. We expect this to be also the case for the exact model.

Identity \eqref{b1} alongside with \eqref{f2} implies
that the spectrum of the exact theory is discrete
with finite multiplicities. As we mentioned above, the
same inequality was demonstrated to be valid for the regularised bosonic model.


\section{On the large $N$ limit of the semiclassical bosonic Hamiltonian}
\label{s4} 

The semiclassical Hamiltonian  of the regularised model
is,
\begin{align*}
H_{\mathrm{sc},N}&=\mathrm{Tr}\,\left[\frac{1}{2N^{3}}
\left(
(P^{m})^{2}+ (\Pi_{r})^{2}\right)+ \frac
{n^{2}}{8\pi^{2}N^{3}}\left(\frac{i}{N}[T_{V_{r}},X^{m}]T_{-V_{r}}
\right)^{2}\right.\\
\nonumber
& +\frac{n^{2}}{16\pi^{2}N^{3}}\left(\frac{i}{N}\right)^{2}([T_{V_{r}},A_{r}]T_{-V_{s}}-
[T_{V_{s}},A_{r}]T_{-V_{r}})^{2} \\ \nonumber
&\left.-\frac{i}{N}\left(\frac{in}{4\pi
N^{3}}\right)\overline{\Psi}\gamma_{-}\gamma_{r}[T_{V_{r}},\Psi]T_{-V_{r}}\right].
\end{align*}
Here the $SU(N)$ matrices
$T_{A}=N\omega^{\frac{1}{2}a_{1}a_{2}}P^{a_{1}}Q^{a_{2}}$,
$T_{0}= N \mathbb{I}$, $A=(a_{1},a_{2})$ and $P,Q$ are the Heisenberg
matrices satisfying the Weyl condition $PQ=\omega QP$ where
$\omega=e^\frac{2\pi i}{N}$. The generators of the algebra of $SU(N)$
may be expressed in terms of $T_{A}$. The fields may be expanded in
this basis as
\( X^{m}=X^{mA}T_{A},\) satisfying
\begin{gather*} [T_{V_{r}},T_{A}]=f_{V_{r}A}^{V_{r}+A}T_{V_{r}+A}\\
T_{A}T_{B}=N\omega^{\frac{1}{2}A\wedge B}T_{A+B}.
 \end{gather*}
Here $f_{A B}^{C}= 2iN\sin(\frac{(A\wedge B)\pi}{N})\delta(A+B-C)$
 are the standard structure constants for the regularisation via $SU(N)$ of a  two-torus, $A\wedge B=
a_{1}b_{2}-a_{2}b_{1}$, $V_{1}=(1,0)$ and $V_{2}=(0,1)$.

The bosonic contribution in $H_{\mathrm{sc},N}$ consists
of two components, one related to the
transverse field sector and the other related to the induced gauge fields on the world-volume of the membrane.

A direct calculation shows that
the transverse field contribution,
\begin{equation*}
\mathrm{Tr}\,\left(\frac{n^{2}}{8\pi^{2}N^{3}}\left(\frac{i}{N}[T_{V_{r}},X^{m}]T_{-V_{r}}\right)^{2}\right),\end{equation*}
is equivalent to
\begin{equation} \label{b3}
V_{\mathrm{sc}_{1},N}=\sum_{A}|X^{mA}|^{2}w_{A,N}^{2}
 \end{equation}
where
\begin{equation*}
w_{A,N}=\frac{nN}{\sqrt{2}\pi}\sqrt{\left(\sin\left(\frac{a_{1}\pi}{N}\right)\right)^2
+\left(\sin\left(\frac{a_{2}\pi}{N}\right)\right)^2}.\end{equation*}

The second contribution corresponds to the gauge fields defined on the
world volume of the membrane as a result of the central
charges induced by the winding.  This contribution is
 \begin{equation}\label{eq1}
V_{\mathrm{sc}_{2},N}=\frac{n^{2}}{8\pi^{2}N^{5}}
\mathrm{Tr}\,(([T_{V_{1}},A_{2}]T_{-V_{1}}-
[T_{V_{2}},A_{1}]T_{-V_{2}})^{2}).
 \end{equation}
Here we fix the
remaining gauge fields by imposing the following constraint
 \begin{equation*}
A_{1}^{(m,n)}=0\quad \mathrm{for}\quad n\neq 0,\qquad A_{2}^{(p,q)}=0
\quad \mathrm{for}\quad q=0.
 \end{equation*}
Then
\begin{equation*}
[T_{V_{1}},A_{2}]T_{-V_{1}}=N\sum_{B}f_{V_{1}B}^{V_{1}+B}\omega^{+\frac{1}{2}b_{2}}T_{B},
\end{equation*}
for $r=1$
and a similar identity holds true for $r=2$.
After performing the canonical reduction associated to the above gauge
fixing conditions, we obtain
 \begin{equation} \label{b2}
V_{\mathrm{sc}_{2},N}=\sum_{B}A_{1}^{B}\overline{A_{1}^{B}}(w_{1,B,N})^{2}+
A_{2}^{B}\overline{A_{2}^{B}}(w_{2,B,N})^{2},
 \end{equation}
 where
\begin{equation} \label{f4}
\begin{aligned}
w_{1,B,N}&=\frac{nN}{\sqrt{2\pi}}\sin\left(\frac{b_{1}\pi}{N}\right)\quad
\mathrm{and}  \\
w_{2,B,N}& =\frac{nN}{\sqrt{2\pi}}\sqrt{\sin\left(\frac{b_{1}\pi}{N}\right)^{2}+
\sin\left(\frac{b_{2}\pi}{N}\right)^{2}}.
\end{aligned}
\end{equation}
These coefficients correspond to contributions as for
the $X^{m}$ modes.

\medskip

By virtue of \eqref{b3} and \eqref{b2}, the bosonic regularised
semiclassical Hamiltonian realises as
\begin{equation} \label{f3}
    H^\mathrm{B}_{\mathrm{sc},N}=-\Delta_{(X^{mA},A_1^B,A_2^B)} +
    V_{\mathrm{sc}_{1},N}+V_{\mathrm{sc}_{1},N}+ c_N,
\end{equation}
a quantum mechanical harmonic oscillator acting on the
Hilbert space $L^2((X^{mA},A_1^B,A_2^B)\in \mathbb{R}^\Lambda)$.
Here $N$ is a large
parameter representing the number of $D0$ branes in the
regularisation process; $(X^{mA},A_1^B,A_2^B)\in \mathbb{R}^\Lambda$
lies in
the space coordinates and $\Lambda=8(N^2-1)$ for an
$SU(N)$ regularisation. Here $A=(m,n)$ and $B=(p,q)$.
The constant $c_N$
is a shift in the position of the ground state energy and it is
chosen, so that the ground energy of
$H^{\mathrm{B}}_{\mathrm{sc},N}$ is exactly zero.

This characterisation of $H^{\mathrm{B}}_{\mathrm{sc},N}$ as an
elliptic partial differential operator is convenient for determining  spectral
\cite{bgmr} and heat kernel \cite{l1} properties of the Hamiltonian of the
supermembrane, once
the regularisation process has been carried out. We now consider
this representation in order to study the large $N$ limit of
$H^{\mathrm{B}}_{\mathrm{sc},N}$ and its connection with
$H^{\mathrm{B}}_{\mathrm{sc}}$. All limiting process below refer to
taking $N\to\infty$.

\begin{lemma}
Each eigenvalue of $H^B_{\mathrm{sc},N}$ converges to a
corresponding eigenvalue of $H^B_{\mathrm{sc}}$ as $N\to \infty$.\end{lemma} The large $N$ limit is defined by taking a fixed energy $E$
and then comparing the spectrum of the operators below $E$ in the limit $N\to\infty$.

\proof Firstly note that there is a one to one correspondence
between individual excited state eigenvalues of
$H^B_{\mathrm{sc,N}}$ and $H^B_{\mathrm{sc}}$, and finite subsets
of $\mathbb{N} \times \mathbb{N}$. When $N<\infty$, the variables
in \eqref{f3} can be separated, so the eigenfunctions of
$H^{\mathrm{B}}_{\mathrm{sc},N}$ are given explicitly in terms of
creation operators as,
\begin{equation*}
a_{C_1}^{(N)\dag}\dots a_{C_{\small \Lambda}}^{(N)\dag}\vert
0\rangle
\end{equation*}
with associated  eigenvalue
\begin{equation*}
\lambda_{\mathcal{F},N}= \omega_{C_1,N}+\dots+\omega_{C_{\small
\Lambda},N}
\end{equation*}
corresponding to the set
\begin{equation*}
   \mathcal{F}=\{C_1,\ldots,C_{\small \Lambda}\}.
\end{equation*}
Here $\omega_{C,N}$ are either $w_{A,N}$, $w_{1,B,N}$ or
$w_{2,B,N}$.
This provides an indexing for the spectrum of
$H^{\mathrm{B}}_{\mathrm{sc},N}$ in terms of finite subsets of
$\mathbb{N}\times\mathbb{N}$ with at most $\Lambda$ elements.
Similarly, for the case of the limiting
$H^{\mathrm{B}}_{\mathrm{sc}}$, the eigenfunctions are constructed
in terms of creation operators, but now the sequences can be of
arbitrary length.  Thus, the eigenvalues are in one to one
correspondence now with all finite subsets of $\mathbb{N}\times
\mathbb{N}$.

For any given finite subset $\mathcal{F}$ of $\mathbb{N} \times
\mathbb{N}$, we just have to choose $N$ larger than the number of
elements of $\mathcal{F}$, in order to ensure that $\mathcal{F}$ is
also included in the indexing for the eigenvalues of
$H^B_{\mathrm{sc},N}$. As $\mathcal{F}$ is finite,
$\omega_{C,N}\to \omega_C$ and the expressions for the eigenvalues
are finite sums,
\begin{equation*}
\lambda_{\mathcal{F},N} \to \lambda_{\mathcal{F}},
\end{equation*}
as required.

\medskip

The following remark on the multiplicity of the
spectrum of $H^B_{\mathrm{sc}}$ is in place.
For a given index $A=(m,n)$, see
\eqref{f2},
\begin{equation*}
   \omega_A\geq \pi^2 \min\{R^1l^1,R^2l^2\},
\end{equation*}
where the constants on the right hand side are independent of $A$.
Then, for a given finite subset $\mathcal{F}$ with $\Phi$
elements,
\begin{equation*}
\lambda_\mathcal{F} \geq \Phi \pi^2 \min\{R^1l^1,R^2l^2\}.
\end{equation*}
Hence, the class of subsets $\tilde{\mathcal{F}}$ such that
$\lambda_\mathcal{F}=\lambda_{\tilde{\mathcal{F}}}$, is
limited by the fact
that $\tilde{\mathcal{F}}$ can not have more than
$\lambda_\mathcal{F}/\pi^2\min\{R^1l^1,R^2l^2\}$ elements. This
ensures that each eigenvalue of $H^B_{\mathrm{sc}}$ is of finite
multiplicity.

\smallskip

The above lemma shows that the spectra of
$H^{\mathrm{B}}_{\mathrm{\mathrm{sc}},N}$ converge to the spectrum
of $H^{\mathrm{B}}_{\mathrm{\mathrm{sc}}}$. However it does not
provide information about the large $N$ limit of
$H^B_{\mathrm{sc},N}$ and its relation with $H^B_{\mathrm{sc}}$.

The Hamiltonian
$H^B_{\mathrm{sc}}$ is equivalent, under unitary transformation,
to a self-adjoint
operator acting on the Hilbert space $L^2(\ell_2,\ud \gamma)$, where
$\ud \gamma$ is a Gaussian measure on $\ell_2$. Recall that $\ell_2$
comprises square summable $(Y^C)_{C\in \N\times\N}$
such that
$\|Y\|^2=\sum_C |Y^C|^2<\infty$. A procedure for constructing
Gaussian  measures in $\ell_2$ is described in the monograph
\cite{kuo}.

For each wave function $\psi \in L^2(\ell_2,\ud \gamma)$, there
exists $\psi_N\in  L^2(\mathbb{R}^\Lambda,\ud \gamma^\Lambda)$
such that $\psi_N\to \psi$. Here $\ud
\gamma^\Lambda=e^{-\|Y\|^2/2} \ud Y$ is the standard Gaussian
measure in $\mathbb{R}^\Lambda$. By performing a suitable change
of coordinates, the Hamiltonian
$H^{\mathrm{B}}_{\mathrm{\mathrm{sc}},N}$ is also an operator
acting on  $L^2(\mathbb{R}^\Lambda,\ud \gamma^\Lambda)$. Hence, we
are in the position of being able to compare the exact model with
the regularised one. Indeed, since $L^2(\mathbb{R}^\Lambda,\ud
\gamma^\Lambda)$ are subspaces of $L^2(\ell_2,\ud \gamma)$ via the
natural identification
\begin{equation*}
    \phi((Y^C)) \longmapsto
    \phi((Y^C),0,\ldots), \qquad (Y^C)\in \R^\Lambda,
\end{equation*}
both operators $H^{\mathrm{B}}_{\mathrm{\mathrm{sc}},N}$ and
$H^{\mathrm{B}}_{\mathrm{\mathrm{sc}}}$ act in the same
subspace. Note that here we must use the fact that $\ud
\gamma$ is Gaussian in order to ensure that the right hand side is
a member of the latter space. In the other direction, we have the
projected states
\begin{equation*}
   \phi(Y^{(1,0)},\ldots) \longmapsto \phi((Y^C),0,\ldots)=:
\phi_N(Y^{(1,0)},\ldots).
\end{equation*}
for all $\phi(Y^{(1,0)},\ldots)\in L^2(\ell_2,\ud \gamma)$. This
identification gives a precise meaning to the limit
\begin{equation*}
    \lim _{N\to \infty} \langle\psi, (H^{\mathrm{B}}_{\mathrm{sc},N}\phi_N -H^{\mathrm{B}}_{\mathrm{sc}} \phi) \rangle,
\end{equation*}
making it possible to verify rigorously whether
$H^{\mathrm{B}}_{\mathrm{sc},N}\to H^{\mathrm{B}}_{\mathrm{sc}}$
in the weak topology.

For given initial states $\phi,\psi\in L^2(\ell_2,\ud \gamma)$
and $t>0$, we can also compute the limit
\begin{equation*}
    \lim_{N\to \infty} \langle \phi_N,e^{-H^{\mathrm{B}}_{\mathrm{sc},N}t} \psi_N\rangle
\end{equation*}
via the Mehler formula. For this we should recall that the heat
kernel of the regularised semiclassical Hamiltonian can be found
explicitly,
$e^{-H^{\mathrm{B}}_{\mathrm{sc},N}t}$ is the
Ornstein-Uhlenbeck semigroup. This rises the question of whether the exact
semigroup $e^{-H^{\mathrm{B}}_{\mathrm{sc}}t}$ could possibly be
characterised using the Feynman-Kac formula.

Note that the characterisation of the regularised Hamiltonian in
the space with Gaussian measure is far more advantageous than our
previous approach of using the space with Lebesgue measure. Indeed
one can easily prove that it is not possible to construct a
Lebesgue measure in $\ell_2$.


\section{Centre of the group, mass gap and confinement}
\label{s6} 

In the first part of this paper we have discussed spectral properties of the supermembrane with central charges.
Let us now investigate its behaviour, or in equivalent manner, the
behaviour of the symplectic NCSYM theory, at extreme energies. This case corresponds to analysing the supermembrane with central charges for $g=3$, that is compactified on $T^{6}$, and performing a further compactification of the fifth coordinate to $S^{1}$. The spectral properties discussed above remain invariant and the theory corresponds to an $N=1$ NCSYM of dimension $2+1$ on a $4-\mathrm{D}$ target-space.

By virtue of the results of \cite{thooft} and \cite{thooft2}, a permanent
quark confinement occurs in a gauge theory, if its vacuum condenses
into a state which resembles a superconductor. This proposal consists
in taking the confinement of quarks as dual to the Meissner effect,
where the roles of magnetism and electricity is interchanged. In this
approach, a non-Abelian gauge theory is considered in terms of
an Abelian theory enriched with Dirac magnetic monopoles. 
As we will see next, this is in
correspondence with the situation we are presently discussing in the case of the $4-\mathrm{D}$ target space, under the  assumption that the underlying $N=1$ Yang-Mills  theory
is symplectic and non-commutative.

Our first task will be to
study the symmetries of the theory. We will then
identify the centre of the group relative to the residual symmetry.
This latter symmetry plays a role
in the creation of confinement and, through
its breaking, the theory enters in a quark-gluon plasma phase
corresponding to the supermembrane without central charges.

\subsection{Symmetries}
\label{s6.1}
The $D=11$ supermembrane in the light cone gauge with a
Minkowski target space possesses a residual invariance, associated to
the infinite group of area preserving diffeomorphisms
$\mathbf{Diff}^{\infty}(\Sigma)$ on a Riemann surface $\Sigma$ of
genus $g$, \cite{dwmn}. The first class constraints of the theory generates
area preserving diffeomorphisms homotopic to the identity,
 $\mathbf{Diff}^{\infty}_{\mathbb{I}}(\Sigma)$.
The local term generates area preserving diffeomorphisms associated to
 exact 1-forms,
 \cite{dwmn}:
\begin{align*}
\partial_{r}(\sqrt{\omega(\sigma)}\xi^{rs}(\sigma,\tau))\equiv
D_{r}\xi^{r}(\sigma)=0 ; \qquad
\quad\xi^{r}(\sigma)=\frac{\epsilon^{rs}}{\sqrt{\omega(\sigma)}}
\partial_{s}\xi(\sigma).
\end{align*}
The global term, on the other hand, generates area preserving
diffeomorphisms associated to harmonic one-forms. The local constraint
corresponds to the Gauss law for the non-commutative symplectic Yang-Mills
formulation. We might call it, the generator of gauge transformations.

Let us analyse the diffeomorphisms generated by the global constraint.
Since only derivatives terms are present,
the harmonic functions on the theory may be defined up to a constant.
Thus, the global constraint
generates an abelian sub-group of diffeomorphisms. In fact, from the
considerations of Section~\ref{s2}, we obtain
\begin{align*}
   \{\xi_{1},\xi_{2}\}=(\mathrm{constant}),
\end{align*}
where $\xi_1$ and $\xi_2$ are harmonic parameters. Hence the commutator
of two such a diffeomorphisms is in the harmonic class of zero. In addition,
each harmonic one-form may be expanded in terms of the basis
$\ud \widehat{X}^r$ as
\[
    \ud \xi =a_1\ud \widehat{X}^1+a_2\ud \widehat{X}^2
\]
where $a_1$ and $a_2$ are real numbers. The algebra of infinitesimal
diffeomorphisms generated by the global constraint contains then a realisation
of $u(1)\times u(1)$. This symmetry of the supermembrane with central charges
preserves gauge equivalent classes. The present
interpretation of this new symmetry is completely different from the
approach followed in \cite{dwpp}, where a $SU(N)$ regularisation of the
supermembrane on a compactified target space was discussed. Moreover,
since a consistent $SU(N)$ regularisation of the supermembrane with central
charges was found in \cite{gmr}, this symmetry should reduce to $\mathbb{Z}_N
\times \mathbb{Z}_N$ in the regularised formulation.

 The presence of $r$ closed, but not exact,
forms can be regarded in the dual picture, i.e. in the $10$-D IIA
description, as the existence of $r$ $U(1)$ gauge fields corresponding
to the compactification process. This means that the compactified
supermembrane has the following gauge symmetries from the type IIA point
of view,
\begin{align*}
\mathbf{Diff}_{\mathbb{I}}^{\infty}\times U(1)^{r}.
\end{align*}

Furthermore, the gauge fields satisfy an additional symmetry
associated to the harmonic forms. The Hamiltonian exhibits an additional
invariance, related to the symplectomorphism group
$\mathrm{Sp}(2g,\mathbb{Z})$. In the particular case of the compactified
sector of target space being a 2-torus, the symmetry becomes
$\mathrm{Sp}(2,\mathbb{Z})\simeq SL(2,\mathbb{Z})$. This is the same symmetry that
arises by compactifying $IIA/S^{1}$ or, equivalently, $IIB/S^{1}$.

As it was pointed
out in \cite{br}, one way to realise this symmetry in our formalism is to observe that $\ud X^{r}$ must satisfy, see (\ref{e2}),
\begin{align*}
\oint_{C_{s}}\ud X^{r}=S_{s}^{r},
\end{align*}
where $C_{s}$ is a basis of homology defined on the 2-torus and
the matrix $S_{s}^{r}$ are such that
\begin{align*}
S_{r}^{t}\epsilon^{rs}S_{s}^{u}=\epsilon^{tu},
\end{align*} when $n=1$ 
i.e. $S_{r}^{s}\in SL(2,\mathbb{Z})$.

In the supermembrane with fixed central charges we find area preserving diffeomorphisms which are not homotopic to the identity. These
diffeomorphisms correspond to bi-holomorphisms, mapping the Teichmuller
space into itself. Under this
conformal mapping, the basis of harmonic one-forms transforms
via an element of  $SL(2,\mathbb{Z})$.
As we saw in Section~{\ref{s2}}, these are area preserving
diffeomorphisms for our choice of $W$.

 Although $A$, cf. (\ref{ee}), is single-valued over $\Sigma$, it
has an infinitesimal gauge transformation law that represents
an unusual realisation of the diffeomorphisms algebra,
\begin{align*}
A \to A+D\xi+\{A,\xi\}=A+\mathcal{D}\xi.
\end{align*}
This transformation is generated by a first class constraint at
both the exact and the regularised level.
See the Appendix. It corresponds to a
symplectic connection, preserving the symplectic structure of the
fibres under holonomies.

With this transformation, the general
structure of the first class constraint which generates the gauge symmetry
of the theory is that of an algebra at both the exact as well as the $SU(N)$
\textit{regularised} model.
We remark that the transversal modes transform in the standard way,
\begin{align*}
\delta X=\{\xi,X\}.
\end{align*}
 In order to construct the
non-commutative gauge theory, one has to fix the harmonic sector.
The resulting symmetry is the centre of $\mathrm{Sp}(2,\mathbb{Z})$
which is $\mathbb{Z}(2)$.

\subsection{The centre of the group as a
mechanism for confinement in the exact theory.}
\label{s6.2}
The mass contribution of the central charge, or, analogously,
its correlated residual $\mathbb{Z}{(2)}$ symmetry of the 
Hamiltonian\footnote{{Strictly speaking the residual symmetry is a $\mathbb{Z}{(2)}\times \mathbb{Z}{(2)}$ 
at classical level, however at quantum level one of the $\mathbb{Z}{(2)}$ symmetries 
is not preserved by the measure of integration
\begin{align}
\det([\Phi,\chi]_{\rm Poisson})
\end{align}
where $\Phi=\mathcal{D}_{r}\Pi^{r}+\{P_{m},X^{m}\}$ and
$\chi=D_{2}A_{1}$. The condition
$\Phi=0$ corresponds to the first class constraint
 and $\chi=0$ corresponds to the gauge fixing condition. The
computation of $[\Phi,X]_{\rm Poisson}=\mathcal{D}_{1}D_{2}$ does not leave this
operator invariant, leaving only the center of the group $\mathbb{Z}{(2)}$ as the residual symmetry.}}, can
be described in terms of the quadratic derivatives of the
configuration fields $X^{m}$ and $A_{r}$.
These are induced
 by the minimal immersion realised by $\widehat{X}_{r}$, $r=1,2$,
the harmonic fields over $\Sigma$,
\begin{align*}
D_{r}Y_{A}=\{\widehat{X}_{r},Y_{A}\}=\lambda_{rA}^{B}Y_{B}=\lambda_{rA}Y_{A}
\end{align*}
where
\begin{align*}
\lambda_{rA}^{B}=\int
d^{2}\sigma\sqrt{\omega}\{\widehat{X}_{r},Y_{A}\}Y^{B}.
\end{align*}
They correspond to a particular
subset of the structure constants that mixes the harmonic and the exact forms,
$g_{rA}^{C}$. For the case of a torus, an explicit relation was found
in \cite{gmr}.
The quadratic terms on 
the derivatives of the configuration variables, define a strictly positive function whose contribution 
to the overall Hamiltonian gives rise to a basin shaped potential. The latter eliminates the string-like 
spikes and provides a discrete spectrum, even for the supersymmetric model.

Without the central charge, the SUSY contribution renders a potential unbounded from below along the 
directions where the quartic contribution vanishes. This would produce a non-empty continuous spectrum.
The quantum mass is bounded
from below by the semiclassical term, see \eqref{fi3} and \eqref{b1}.
Then, once the topological condition is
implemented, the centre
created by a discrete symmetry is a mechanism for rendering mass to the
monopoles.

A natural question to ask is what happens
when we compactify the target space by a larger space, say $T^{6}$.
The size of the symplectic group
increases. However, as explained in \cite{pepe2}, the size of the
centre of the group remains constant modulo $\mathbb{Z}(2)$, in distinction to
the $SU(N)$ gauge groups. From a lattice point of view,
$\mathrm{Sp}(2)$  and $\mathrm{Sp}(3)$ were used to demonstrate de-confined
transition phases induced by the breaking of the centre.

\subsection{Centre of the group as a mechanism for confinement in the $SU(N)$ formalism}
\label{s6.3}
The centre of the group in the $SU(N)$ regularisation is known to be
$\mathbb{Z}_{N}$. Since the origin of this symmetry is an inherited
structure of topological nature created by the monopoles induced in
the torus, the real discrete symmetry is $\mathbb{Z}_{N}\times
\mathbb{Z}_{N}$. The realisation of the latter in terms of the field
theory should correspond to a symmetry-preserving
gauge equivalent classes. Although we do not have the explicit form of
such a map, we argue that it should exist and it should be related
to the mass terms. In fact, the exact theory has a symmetry
$U(1)\times U(1)$, also in the large $N$ limit
\[
    \mathbb{Z}_N \times \mathbb{Z}_N \to U(1)\times U(1),
\]
and ultimately the large $N$ limit of the regularised supermembrane
with central charges indeed corresponds to the exact theory. Consequently,
the former should have that symmetry.
An element $\widehat{z}_{rA}\in \mathbb{Z}_{N}\times \mathbb{Z}_{N}$ belongs to the centre, if it satisfies
the condition
\begin{align*}
 \widehat{z}_{rA}^{N}=1,\qquad \widehat{z}_{rA}=e^{\frac{2\pi
i(V_{r}\wedge A)}{N}}.
\end{align*}

The contributions associated to the mass
are defined in terms of a regularised object found in
\cite{gmr}. They correspond to a specific choice of the structure
constants. In terms of the $SU(N)$ basis,
\begin{align*}
\widehat{\lambda}_{rA}\equiv
\mathrm{Tr}([T_{V_{r}},T_{A}]T_{-V_{r}-A})=2iN\sin\left(\frac{V_{r}\wedge
A}{N}\pi\right),
\end{align*}
where $T_{V_{r}}$ correspond to two particular matrices
of $T_{A}$ from which the $SU(N)$ algebra can be expanded. 
The centre is \begin{align*}
\widehat{z}_{rA}\equiv\frac{1}{N^{4}}\mathrm{Tr}(T_{V_{r}}T_{A}T_{-V_{r-A}}).
\end{align*}
Following \cite{gmr}, these matrices are given by
$T_{V_{1}}=T_{0,1}$ and $T_{V_{2}}=T_{1,0}$. Here we
have used that $T_{A}=N z^{1/2
a_{1}a_{2}}P^{a_{1}}Q^{a_{2}}$ and $T_{(0,0)}=N\mathbb{I}$ as in \cite{dwhn,thooft}.
The lambda contribution can then be re-written as
\begin{align*}
\widehat{\lambda}_{rA}=\mathrm{Im}(\widehat{z}_{rA}).
\end{align*}
The discrete mass spectrum arises in analogous way as in
the exact theory.

 The
eigenvalues of the Hamiltonian are bounded from below by those of
the semiclassical spectrum in such a way, that the mass terms are
created by the centre. The unitary realisation of this centre in the Hilbert space of states,
commutes with the Hamiltonian and it therefore represents an unbroken symmetry.

\subsection{Confinement, screening and phase transition}
\label{s6.4}
As we have already seen, the mass terms are determined by the
elements of the centre  $m(\widehat{z}_{rA})$ associated to
$\widehat{z}_{rA}$. The matrix $T_{V_{r}}$
appears in the regularised model as a consequence of the presence of the term
$\widehat{X}_{r}$. The latter are the
harmonic forms associated to the winding, defining the monopole charge
$\{\widehat{X}_{r},\widehat{X}_{s}\}=\epsilon_{rs}n$.
If the monopole charge disappears, the centre becomes trivial
$\widehat{z_{A}}=N^{-3}Tr({T_{A}T_{-A}})=N^{-2}Tr({T_{(0,0)}})=1$. We therefore expect a breaking of the
centre of the group for the
de-confined phase. In a similar fashion, the same effect
can be seen at the level of the exact theory.

We have then two pictures. One, corresponding to the supermembrane with
central charges, in which the correlation length of
the particles is the inverse of the mass of the glueball states,
$\xi_{C}=1/m_{\mathrm{eff}}$, and we can define an effective volume
$V_{\mathrm{eff}}=R/\xi_{C}$. There the particles feel the topological
effects and get confined. The other one in which $m^{2}=0$ as
$\xi\to\infty$ and $V_{\mathrm{eff}}=0$. There
the particles lose the information
that they are confined in a boundary with topological condition, and
behave as if they where in a quark-gluon plasma.

In the supersymmetric picture, the
Hamiltonian corresponding to the $N=1$ supermembrane with central charges,
\begin{align*}
&H_{1}=\Delta_{1}+V_{1}^{B}+V^{F}_{1}.
\end{align*}
It has purely discrete spectrum at the quantum level, due to the presence
of a non-trivial central charge in the algebra of supersymmetry.

 In contrast, the
Hamiltonian corresponding to the $N=4$ compactified supermembrane
without central charges,
\begin{align*}
&H_{2}=\Delta_{2}+V_{2}^{B}+V^{F}_{2},
\end{align*}
has a continuum spectrum. The supermembrane is interpreted as a
many-body object fluctuating into different vacua were neither the
number of particles nor the topology of the membrane is
preserved \cite{dwln,helling}. We conjecture that this case describes the
quark-gluon plasma.

Moreover, since the potential vanishes, the particles are free
along the flat directions corresponding to the commutative picture.
The particles do not feel any force between them, as it is expected in the
asymptotic free regime of a SUSY QCD. The transition is then a consequence
of the presence of a quantum change in the irreducible winding.
Although both  types of membranes are compact and can have the same topology,
there is a change of quantum nature
in the topological condition. It corresponds to the sequence
\begin{align*}
   U(1)\times U(1)\longrightarrow U(1)\longrightarrow \mathbb{Z}(2),
\end{align*}
related to a monopole bounding two strings.

We
conjecture about the origin of this phenomenon.
If we decrease the energy scale, it becomes more
advantageous for the membrane to have an irreducible wrapping.
Since the radius of the compactified
extra dimension becomes smaller as we lower the scale, there is a
critical scale at which the area of the supermembrane is minimised,
not by wrapping in a cycle, but by considering
a calibrated submanifold
generated by the  monopoles dual to the irreducible wrapping.
For an account on the quantum topological change in (2+1)D see
\cite{balachandran,preskill}.

\subsection{Supermembrane origin and interpretation of SUSY QCD}
\label{s6.5}

We would like to stress that, in our picture,
confinement is a consequence
of two issues: supersymmetry and the addition of
extra dimensions. By virtue of the topological condition imposed
on the extra dimensions, the
supersymmetry is broken at critical energies.
We believe that this topological condition
appears naturally when the size of the extra dimensions decrease.
It corresponds to
the presence of a central charge of the supersymmetric algebra and
it adds mass to the gluons entering a confined phase.

Since the
magnetic flux is confined in the monopoles picture (as it was
originally explained in \cite{thooft2}), the electric flux between them
gets also confined, forming a $Z_{2}$-string at the ends of which
quarks become attached \cite{kneipp}. In order to separate them, one
needs to provide a force proportional to the force needed
for increasing the effective radius of the compact dimensions. This force
grows linearly with the radius. In this picture the confinement of quarks
is then consequence of the compactification of extra dimensions
and the fact that
supermembrane has an irreducible wrapping around them.

Separating the
gluons means de-compactifying the space. In higher energies, the size
of the effective radius of the extra dimensions becomes bigger. At those energies the supermembrane does not minimise its energy with an
irreducible wrapping around  it (which corresponds to wrap a
calibrated submanifold), but just wraps in cycles that minimise
their volume.

It is known that the presence of topological defects
can diminish the energy of the vacuum. This is what happens in
our case.  Without these defects, the compactification process
allows degenerate points on the metric. Changes in the
metric and the topology are also allowed in the classical analysis of GR
\cite{horowitz}. This issue has been studied in
\cite{balachandran, preskill,kronfeld}.

A change in the
topology \cite{balachandran} leads then to a loss of the monopole picture.
The centre of the group becomes trivial, hence the theory
enters in a quark-gluon plasma phase. In the latter state,
the supermembrane
can not be associated to a single particle but rather
to a multiple body object in a potential of continuous spectrum. The theory contains free quarks  along the commutative directions that corresponds to a vanishing potential.

Since the correlation length becomes
infinite and the effective volume is zero,
the quarks-gluons form a plasma that does not
feel the boundary effects.
This corresponds to seing inside the hadron, i.e.
at a shorter scale.

A further natural explanation emerges from here.
Supersymmetry is where the topological condition originates.
We conjecture that, perhaps, this is the way in which
supersymmetry breaking realises. This type of compactification
does not produce exotic matter as in the KK
reduction, but rather gives mass (without a Higss mechanism)
to the scalar fields as the supersymmetry is broken.
We may speculate that supersymmetry, membrane description and the addition of
extra dimensions, would be the reason for a QCD behaviour in both
phases:  the confined one and the quark-gluon plasma one.

\section{Discussion and conclusions} \label{s7}
Our results provide bounds, in the operator sense, for the bosonic
Hamiltonian of the $D=11$ supermembrane with central charges. We have shown
that this
Hamiltonian is bounded from below by a strictly positive constant,
times the Hamiltonian of coupled harmonic oscillators.
We have then shown
that the spectrum of the Hamiltonian is discrete with finite
multiplicity and it contains a mass gap. This result extends to the exact
theory the bound already obtained for
the $SU(N)$ regularised model. We are not aware of any other
similar analysis in the literature,
regarding the spectrum of the exact supermembrane theory rather
than that of the $SU(N)$ regularised theory.
Our bound gives strong evidence that the quantum properties of the supersymmetric 
potential will not change qualitatively those of the bosonic potential. 

In previous works we have analysed the heat kernel of
the regularised supermembrane with central charges.
Convergence of the kernel's asymptotic expansion
was obtained in the Schatten-von Neumann norms,
implying a well defined Feynman formula. The
large $N$ limit of such formula is expected to converge to the Feynman
integral of the supermembrane with central charges.

Since this
theory is the quantum equivalent of a symplectic NCSYM
theory in $2+1$ dimensions, the same spectral properties are valid for
the latter. The symplectic NCSYM in $2+1$ dimensions
is coupled to scalar
fields arising from the dimensional reduction of
NCYM in $9+1$ dimensions.
The number of degrees of freedom of both theories coincides.
In this context, our results provide a one-step-forward
towards the quantisation of M-theory.

We have also argued that the supermembrane theory,
when compactified in $4$-D, can
be interpreted as a theory modelling SUSY QCD. As the theory
becomes the $N=1$ supermembrane with central charges
at zero temperatures, it exhibits
confinement in the phase. By rising
the energy, the theory enter in a phase of asymptotic freedom
described by the $N=4$ compactified supermembrane without central charges.
We have shown that the phase transition is described by
the breaking of the centre of the group.

We have also conjectured a possible reason for this phase transition.
At high energies, the size of the effective radius of the extra
dimensions becomes bigger. In the same regime,
the irreducible wrapping of the
supermembrane on a calibrated submanifold of the target, turns into a
reducible wrapping in the compact sector of the target space with
zero central charge.
This corresponds to seeing inside the hadron, that
is at a shorter scale. Since the correlation length becomes
infinite and the effective volume is zero,
the quarks-gluons form a plasma that does not feel
the topological effects. Along the commutative
directions, the quarks experience no force.


\section{Appendix} \label{a1}
Here we compute explicitly the $SU(N)$ gauge
symmetry in the regularisation of the $D=11$ supermembrane with
central charges.

The general structure of the first class
constraints which generate the gauge symmetry of the regularised
model arising from the $D=11$ supermembrane with central charges
is,
\begin{align*}
\phi^{D}\equiv\lambda
f_{E-w_{s},w_{s}}^{D}\Pi^{Es}+f_{A+w_{s},F-w_{s}}^{D}A_{s}^{A}\Pi^{Fs}=0.
\end{align*}
The constant $\lambda$ is an arbitrary parameter and
$f_{AB}^{C}$ are the $SU(N)$ structure constants.

The algebra
associated to the first class constraints is obtained by
considering the Poisson brackets of the generators. We have,
\begin{align*}
[\lambda
f_{E-w_{s}}^{D},&w_{s}\Pi^{Es}+f_{A+w_{s},F-w_{s}}^{D}A_{s}^{A}\Pi^{Es},
f_{L-w_{r}}^{C},w_{r}\Pi^{Lr}+f_{F+w_{r},E-w_{r}}^{C}A_{r}^{F}\Pi^{Er}]_{P}=\\
 &\lambda
f_{E-w_{s},w_{s}}^{D}f_{-E+w_{s},F-w_{s}}^{C}\Pi^{Fs}-\lambda
f_{-L+w_{s},F-w_{s}}^{D}f_{L-w_{r},w_{r}}^{C}\Pi^{Fs}\\
&+f_{A+w_{s},E-w_{s}}^{D}f_{-E+w_{s},F-w_{s}}^{C}A_{s}^{A}\Pi^{Fs}-f_{E+w_{s},F-w_{s}}^{D}f_{A+w_{s},-E-w_{s}}^{C}A_{s}^{A}\Pi^{Fs}.
\end{align*}
We now use the explicit expression for the structure constants.
For the terms depending on $\lambda$ we obtain
\begin{align*}
\lambda\delta^{D+C}_{F}\Pi^{Fs}N^{2}\sin\left(\frac{(C\wedge
D)}{N}\pi\right)\sin\left(\frac{(D+C)\wedge w_{s}}{N}\pi\right)
\end{align*}
which may be re-written as $\lambda f_{-D,E}^{C}f_{F-w_{s},w_{s}}^{E}\Pi^{Fs}$.
For the remaining terms we obtain the expression
\begin{align*}
N^{2}\delta_{A+F}^{C+D}&\sin\left(\frac{(A+w_{s})\wedge
D}{N}\pi\right)\sin\left(\frac{C\wedge (F-w_{s})}{N}\pi\right)\\
\nonumber &
-\sin\left(\frac{D\wedge (F-w_{s})}{N}\pi\right)\sin\left(\frac{(A+w_{s})\wedge
C)}{N}\pi\right),
\end{align*}
which may be re-written as $f_{-D,E}^{C}f_{A+w_{s},F-w_{s}}^{E}A_{s}^{A}\Pi^{Fs}.$
Thus, for any $\lambda$ and $N$,
\begin{align}\label{I}
[\phi^{D},\phi^{C}]_{\mathrm{Poisson}}=f_{-D,E}^{C}\phi^{E}
\end{align}
where $f$ are the $SU(N)$ structure constants.

\medskip

The analogous computation
is valid as we take the large $N$ limit. In
fact, by considering the same constraints, but with the structure
constants associated to the area preserving diffeomorphisms,
we obtain the corresponding algebra (\ref{I}).

As $N\to\infty$, we find an equivalent realisation of the
generators in terms of the constraints,
\begin{align*}
\widetilde{\phi}^{D}=\lambda
f_{E-w_{s},w_{s}}^{D}\Pi^{Es}+f^{D}_{A,F}A_{s}^{A}\Pi^{Fs}=0.
\end{align*}
They also satisfy
\begin{align*}
[\widetilde{\phi}^{D},\widetilde{\phi}^{C}]_{\mathrm{Poisson}}
=f_{-D,E}^{C}\widetilde{\phi}^{E}.
\end{align*}
As a particular case, we can take $\lambda=1$.

In the
regularised model analysed in \cite{gmr}, we wrote the exact
model in
terms of the decomposition on an orthonormal basis over the
Riemann surface. For this, we fixed the gauge and then obtained a regularised
model. An alternative approach is to obtain the
regularised model satisfying the symmetry generated by $\phi^{D}$
and then  perform the gauge fixing.

\section*{Acknowledgements}
We are very grateful to M.~Asorey, F.~Cachazo, I.~Martin,  H.~Nicolai and
Sh.~Matsuura for helpful discussions and comments. We are also
indebted to C.~Burgess, F.~Cachazo, R.~Myers and F.~Quevedo, for
their support and encouragement. M.P.G.M. and A.R. would like to
thank the Perimeter Institute for its kind hospitality where part of
this work was carried out. M.P.G.M acknowledge support from NSERC
Canada, MEDT Ontario and the NSERC Discovery grants program. The work of A.R. was supported by PROSUL, under contract CNPq 490134/2006-08.

\newpage

\begin{minipage}[c]{12cm}
\noindent{\small
\textsf{\hspace{-.18cm}$^{1}$Department of Mathematics and
the Maxwell Institute for Mathematical
Sciences, Heriot-Watt University,
Edinburgh EH14 2AS, United Kingdom.} \newline
\emph{email:} \texttt{L.Boulton@hw.ac.uk}}
\end{minipage}

\vspace{1cm}

\begin{minipage}[c]{12cm}
\noindent {\small
\textsf{\hspace{-.18cm}$^{2}$Perimeter Institute for Theoretical Physics,
Waterloo, Canada, Ontario N2L 2Y5, Canada. \newline
Dept. of Physics and Astronomy, MacMaster University,
1280 Main Street West, Hamilton, Ontario, L8S 4M1, Canada. \newline
DAMTP, DAMTP, Centre for Mathematical Sciences, University of
Cambridge, Cambridge CB3 0WA, United kingdom.} \newline
\emph{emails:} \texttt{mmoral@perimeterinstitute.ca \newline
\null \hspace{1.15cm} M.G.d.Moral@damtp.cam.ac.uk}}
\end{minipage}

\vspace{1cm}

\begin{minipage}[c]{12cm}
\noindent{\small
\hspace{-.18cm}\textsf{$^{3}$Departamento de F\'\i sica,
Universidad Sim\'on Bol\'\i
var, Apartado 89000, Caracas 1080-A, Venezuela.} \newline
\emph{email:} \texttt{arestu@usb.ve}}
\end{minipage}


\begin{thebibliography}{99}

\bibitem{bst} E. Bergshoeff, E. Sezgin, P.K. Townsend,
{\em Supermembranes and eleven-dimensional supergravity.} Phys.
Lett. {\bf B189}: 75-78, 1987.

\bibitem{dwln} B. de Wit, M. Luscher, H. Nicolai, {\em The supermembrane is unstable}.
Nucl. Phys. {\bf B320}: 135, 1989.

\bibitem{dwmn} B. de Wit, U. Marquard, H. Nicolai,
{\em Area preserving diffeomorphisms and supermembrane lorentz
invariance.} Commun. Math. Phys. {\bf 128}: 39-62, 1990.

\bibitem{dwhn} B. de Wit, J. Hoppe, H. Nicolai, {\em On the quantum mechanics of
supermembranes}. Nucl. Phys. {\bf B305}: 545,1988.

\bibitem{hoppe} J. Hoppe, Ph.D Thesis, MIT, 1982.

\bibitem{halpern} M. Claudson, M. B. Halpern, {\em Supersymmetric
ground state wave functions}. Nucl. Phys.{\bf B250}: 689, 1985.

\bibitem{lucher} M. L\"uscher, {\em Some analytic results concerning the mass spectrum of Yang-Mills
gauge theories on a torus}. Nucl.Phys. {\bf B219}: 233-261, 1983.

\bibitem{amilcar} M.P. Garcia del Moral, L. Navarro,
A.J. Perez, A. Restuccia, {\em Intrinsic moment of inertia of
membranes as bounds for the mass gap of Yang-Mills theories.} Nucl.
Phys.{\bf B765}: 287-298, 2007. {\tt hep-th/0607234}

\bibitem{dwpp} B. de Wit, K. Peeters, J. Plefka,
{\em Supermembranes with winding.} Phys. Lett. {\bf B409}: 117-123,
1997. {\it hep-th/9705225}

\bibitem{helling} H Nicolai, R Helling, {\em Supermembranes and matrix theory.} 1998. {\tt hep-th/9809103}

\bibitem{torrealba} I. Martin, A. Restuccia, R. S. Torrealba,
{\em On the stability of compactified D = 11 supermembranes.}
 Nucl. Phys. {\bf B521}: 117-128, 1998. {\tt hep-th/9706090}

\bibitem{gmr} M.P. Garcia del Moral, A. Restuccia, {\em On the
spectrum of a noncommutative formulation of the D=11 supermembrane
with winding} Phys.Rev. D66 045023, 2002. {\tt hep-th/0103261}

\bibitem{bgmmr} L. Boulton, M. P. Garcia del Moral, I.
Martin, A. Restuccia  {\em On the spectrum of a matrix model for the
D=11 supermembrane compactified on a torus with non-trivial
winding.} Class. Quant. Grav. {\bf 19}  2951, 2002. {\tt
hep-th/0109153}

\bibitem{bgmr} L. Boulton, M.P.
Garcia del Moral, A. Restuccia, {\em Discreteness of the spectrum of
the compactified D=11 supermembrane with non-trivial winding}.
Nucl.Phys. {\bf B671} 343-358, 2003. {\tt hep-th/0211047}

\bibitem{l1} L. Boulton and A. Restuccia,
{\em The Heat kernel of the compactified D=11 supermembrane with
non-trivial winding}. Nucl. Phys. {\bf B724} 380-396, 2005. {\tt
hep-th/0405216}

\bibitem{br} J. Bellorin, A. Restuccia,
{\em D=11 Supermembrane wrapped on calibrated submanifolds}
Nucl.Phys. {\bf B737} 190-208, 2006. {\tt hep-th/0510259}

\bibitem{d2-d0} M.P. Garcia del Moral, A. Restuccia,
{\em The Supermembrane with central charge as a bundle of D2 - D0
branes.} Institute of Physics Conference Series 2005, Vol {\bf 43}, 151.
{\tt hep-th/0410288}

\bibitem{rita} R. Gianvittorio, A. Restuccia, J. Stephany,
{\em  Interacting D2-branes in 10 dimensions and non Abelian
Born-Infeld theory.} Class. Quant. Grav. {\bf 23}: 7471-7478, 2006. {\tt
hep-th/0606063}


\bibitem{ovalle}I. Martin, J. Ovalle, A. Restuccia, {\em D-branes, symplectomorphisms and noncommutative gauge theories.}
Nucl. Phys. Proc. Suppl. {\bf 102}: 169-175, 2001; {\em Compactified D =
11 supermembranes and symplectic noncommutative gauge theories. }
Phys. Rev.{\bf D64}: 046001, 2001. {\tt hep-th/0101236}

\bibitem{mr}I. Martin  and A. Restuccia, {\em Symplectic connections,
noncommutative Yang-Mills theory and supermembranes.} Nucl.Phys. {\bf
B622}: 240-256, 2002. {\tt hep-th/0108046}

\bibitem{gross} D. Gross, A. Hashimoto, N. Itzhaki, {\em  Observables of noncommutative gauge theories.}
Adv. Theor. Math. Phys. {\bf 4}: 893-928, 2000. {\tt hep-th/0008075}

\bibitem{szabo} R.J. Szabo, {\em Symmetry, gravity and noncommutativity.}
 Class. Quant. Grav.{\bf 23}: R199-R242, 2006. {\tt hep-th/0606233}

\bibitem{cornalba-schiappa} L. Cornalba, R. Schiappa, {\em Nonassociative star product deformations for D-brane world volumes in curved backgrounds.}
Commun. Math. Phys.{\bf 225}: 33-66, 2002. {\tt hep-th/0101219}


\bibitem{douglas-connes} A. Connes, M. Douglas, A. Schwarz,
 {\em Noncommutative geometry and matrix theory: Compactification on tori.}
 JHEP 9802: 003, 1998. {\tt hep-th/9711162}

\bibitem{miao} Miao Li,
 {\em Comments on supersymmetric Yang-Mills theory on a noncommutative torus.} 1998. {\tt hep-th/9802052}

\bibitem{douglas-hull} M. Douglas, C. Hull,
 {\em D-branes and the noncommutative torus.} JHEP {\bf 9802}: 008,1998. {\tt hep-th/9711165}

\bibitem{ishibashi} N. Ishibashi, {\em A Relation between commutative and noncommutative descriptions of
D-branes.} Beijing Frontiers of theoretical physics, 99-109, 1999,
and
Hayama Noncommutative differential geometry and its
applications to physics 49-61, 1999. {\tt hep-th/9909176}

\bibitem{bigatti} D. Bigatti, L. Susskind, {\em Noncommutative geometry and superYang-Mills theory.}
Phys. Lett. {\bf B451}: 324-335, 1999. {\tt hep-th/9804120}

\bibitem{maldacena-ruso} J. M. Maldacena, J. G. Russo, {\em Large N limit of noncommutative gauge theories.}
JHEP {\bf 9909}: 025, 1999. {\tt hep-th/9908134}

\bibitem{hashimoto} A. Hashimoto, N. Itzhaki, {\em On the hierarchy between noncommutative and
ordinary supersymmetric Yang-Mills.} JHEP {\bf 9912}: 007, 1999. {\tt
hep-th/9911057}

\bibitem{BFSS} T. Banks, W. Fischler, S.H. Shenker, L. Susskind,
 {\em M Theory as a matrix model: a conjecture}.
Phys. Rev. {\bf D55} 5112-5128, 1997. {\tt hep-th/9610043}

\bibitem{bmn} D. Berenstein, J. M. Maldacena, H. Nastase, {\em Strings in flat space
and pp waves from N=4 superYang-Mills.} JHEP {\bf 0204}: 013, 2002. {\tt
hep-th/0202021}

\bibitem{pepe} M. Pepe, {\em Confinement and the center of the gauge group.}
Nucl. Phys. Proc. Suppl. {\bf 153}: 207-214, 2006. {\tt
hep-lat/0510013}

\bibitem{kuo} Kuo Hui Hsiung {\em Integration in Banach space.}
Notes in Banach spaces,  pp. 1--38, Univ. Texas Press, Austin, Tex., 1980.

\bibitem{thooft} G. t'Hooft, {\em On the phase transition towards permanent quark confinement.}
Nucl. Phys. {\bf B138}: 1, 1978.

\bibitem{thooft2} G. 't Hooft, {\em Topology of the gauge condition and new confinement phases in
nonabelian gauge theories.} Nucl. Phys. {\bf B190}: 455, 1981.


\bibitem{aharony} O. Aharony, E. Witten, {\em Anti-de Sitter space and the center of the
gauge group.} JHEP {\bf 9811}: 018, 1998. {\tt hep-th/9807205}

\bibitem{pepe2} K. Holland, M. Pepe, U.J. Wiese, {\em The Deconfinement
phase transition of Sp(2) and Sp(3) Yang-Mills theories in (2+1)-dimensions and (3+1)-dimensions.}
Nucl. Phys. {\bf B694}: 35-58, 2004. {\tt hep-lat/0312022}

\bibitem{witten} E. Witten,
{\em Instantons, the quark model and the 1/N expansion.}\\
 Nucl. Phys. {\bf B149}: 285-320, 1979.

\bibitem{gabadadze} G. Gabadadze, {\em Modeling the glueball spectrum by a closed bosonic
membrane.} Phys. Rev. {\bf D58}: 094015, 1998. {\tt hep-ph/9710402}

\bibitem{chodos} A. Chodos, R.L. Jaffe, K.
Johnson, Charles B. Thorn, V.F. Weisskopf, {\em A new extended model
of hadrons.} Phys. Rev. {\bf D9}: 3471-3495, 1974.

\bibitem{yaffe} B. Svetitsky, L. G. Yaffe, {\em Critical Behavior At Finite Temperature Confinement Transitions.}
Nucl.Phys.{\bf B210}: 423, 1982.


\bibitem{preskill} J.Preskill, A. Vilenkin, {\em Decay of metastable topological defects.}
 Phys. Rev. {\bf D47}: 2324-2342, 1993. {\tt hep-th/9209210}

\bibitem{kronfeld} A. S.
Kronfeld, G. Schierholz, U.J. Wiese, {\em Topology And Dynamics Of
The Confinement Mechanism.} Nucl. Phys. {\bf B293}: 461, 1987.

\bibitem{balachandran} A.P. Balachandran, E. Batista,
I.P. Costa e Silva, P. Teotonio-Sobrinho, {\em Quantum topology
change in (2+1)-dimensions.} Int. J. Mod. Phys. {\bf A15}:
1629-1660, 2000. {\tt hep-th/9905136}

\bibitem{horowitz} G. T. Horowitz, {\em Topology change in classical and quantum
gravity.}  Class. Quant. Grav. {\bf 8}: 587-602, 1991.

\bibitem{stelle} M.J. Duff, T.
Inami, C.N. Pope, E. Sezgin, K.S. Stelle, {\em Semiclassical
Quantization Of The Supermembrane.} Nucl. Phys. {\bf B297}: 515,
1988.

\bibitem{kneipp} Marco A.C. Kneipp, {\em Z(k) string fluxes and monopole confinement in non-Abelian
theories.} Phys. Rev. {\bf D68}: 045009, 2003. {\tt hep-th/0211049}



\end{thebibliography}
\end{document}